\documentclass[twocolumn,superscriptaddress,showpacs,nofootinbib,notitlepage,preprintnumbers,secnumarabic,amssymb, nobibnotes, aps, prl]{revtex4-2}
\usepackage[utf8]{inputenc}
\usepackage{graphicx}
\usepackage{latexsym,amsmath,amssymb,amsthm,lmodern,float,url}
\usepackage{natbib}
\usepackage{color}
\usepackage{microtype}
\usepackage{import}
\usepackage{bbold}
\usepackage[plain]{fancyref}
\usepackage{varioref}
\usepackage{slashed}
\usepackage{multirow}
\usepackage{tikz}
\usepackage{braket}
\usepackage{mathtools}
\usetikzlibrary{shapes}
\usetikzlibrary{positioning}
\usetikzlibrary{decorations.markings}
\usepackage[normalem]{ulem}
\usepackage{qcircuit}
\usepackage{units}
\usepackage{subcaption}

\newcommand{\fig}[1]{Fig.~\ref{fig:#1}}
\newcommand{\tab}[1]{Tab.~\ref{tab:#1}}

\usepackage[colorlinks=true,backref=false, linktocpage=true,
citecolor=blue,urlcolor=blue,linkcolor=blue,pdfpagemode=UseOutlines]{hyperref}
\hypersetup{%
  bookmarksnumbered=true,
  pdftitle = {},
  pdfsubject = {},
  pdfauthor = {},
  pdfkeywords = {}
}

\DeclareMathOperator{\tr}{Tr}
\DeclareMathOperator{\re}{Re}

\tikzset{->-/.style={decoration={
  markings,
  mark=at position .5 with {\arrow{>}}},postaction={decorate}}}

\begin{document}
\preprint{FERMILAB-PUB-21-674-T}
\title{Improved Hamiltonians for Quantum Simulations of Gauge Theories}% Force line breaks with \\
\author{Marcela Carena}
\email{carena@fnal.gov}
\affiliation{Fermi National Accelerator Laboratory, Batavia,  Illinois, 60510, USA}
\affiliation{Enrico Fermi Institute, University of Chicago, Chicago, Illinois, 60637, USA}
\affiliation{Kavli Institute for Cosmological Physics, University of Chicago, Chicago, Illinois, 60637, USA}
\affiliation{Department of Physics, University of Chicago, Chicago, Illinois, 60637, USA}
\author{Henry Lamm}
\email{hlamm@fnal.gov}
\affiliation{Fermi National Accelerator Laboratory, Batavia,  Illinois, 60510, USA}
\author{Ying-Ying Li}
\email[Corresponding author: ]{yingying@fnal.gov}
\affiliation{Fermi National Accelerator Laboratory, Batavia,  Illinois, 60510, USA}
\author{Wanqiang Liu}
\email{wanqiangl@uchicago.edu}
\affiliation{Department of Physics, University of Chicago, Chicago, Illinois, 60637, USA}
\date{\today}

\begin{abstract}
Quantum simulations of lattice gauge theories for the foreseeable future will be hampered by limited resources. The historical success of improved lattice actions in classical simulations strongly suggests that Hamiltonians with improved discretization errors will reduce quantum resources, \textit{i.e.} require $\gtrsim 2^d$ fewer qubits in quantum simulations for lattices with $d$ spatial dimensions. In this work, we consider $\mathcal{O}(a^2)$-improved Hamiltonians for pure gauge theories and design the corresponding quantum circuits for its real-time evolution in terms of primitive gates. An explicit demonstration for $\mathbb{Z}_2$ gauge theory is presented including exploratory tests using the\texttt{
ibm\_perth} device. 
\end{abstract}

\maketitle
\noindent
\textit{Introduction - }
Monte Carlo methods in lattice gauge theory (LGT), though powerful in many nonperturbative calculations, can suffer from sign problems - the Boltzmann weight during sampling becomes complex-valued - when simulating real-time dynamics. Thus, exponential resources are required to solve many interesting problems in particle physics, such as out-of-equilibrium evolution in the early universe~\cite{Yamamoto:2014vda}, 
parton distribution function in hadron collisions \cite{Lamm:2019uyc,Kreshchuk:2020dla,Echevarria:2020wct}, and the shear viscosity of the quark-gluon plasma~\cite{Cohen:2021imf}. 
Quantum computers can directly perform real-time simulations, avoiding these exponentially large resources plaguing classical methods~\cite{Feynman:1981tf,Jordan:2017lea,Banuls:2019bmf}.
Quantum simulation in the Hamiltonian formalism evolves the system with the time evolution operator $ \hat{\mathsf{U}}(t) = e^{-i \hat H t}$. A Hamiltonian $\hat H$ is constructed at finite lattice spacing $a$, causing discretization errors compared to the continuum theory in powers of $a$. 
%depending upon $a$. 
Hamiltonians with discretizations scaling with lower powers of $a$ require smaller lattice spacings for the same errors. This implies larger qubit requirements since the number of qubits is $\mathcal{O}((L/a)^d)$ for a $d$ spatial dimensional lattice of length $L$.

The lattice gauge degrees of freedom, e.g. photons and gluons, need to be rendered finite and mapped to qubits~\cite{digi_loi, Hackett:2018cel,Alexandru:2019nsa,Yamamoto:2020eqi,Ji:2020kjk,Haase:2020kaj,Zohar:2012ay,Zohar:2012xf,Zohar:2013zla,Zohar:2014qma,Zohar:2015hwa,Zohar:2016iic,Klco:2019evd,Ciavarella:2021nmj,Bender:2018rdp,Wiese:2014rla,Luo:2019vmi,Brower:2020huh,Mathis:2020fuo,Singh:2019jog,Singh:2019uwd,Buser:2020uzs}. Current estimates for representing $SU(3)$ suggest $\sim 10$ qubits per gluon link
~\cite{Alexandru:2019nsa,Raychowdhury:2018osk,Raychowdhury:2019iki,Davoudi:2020yln,Ciavarella:2021nmj,Kan:2021xfc,Alexandru:2021jpm}.
%This implies that kiloqubyte-sized computers are required for 3+1d $SU(3)$ simulations even for a small system with $5^3$ lattice size\hank{why $5^3$?}. 
Further exacerbating the demand for qubits is the current, noisy status of quantum computers due to, e.g. entanglement with the environment and imperfect evolution. Though it remains an open question of how much quantum error correction is required to perform lattice simulations, general estimates suggest $\mathcal{O}(10^{1-5})$ physical qubits per logical qubit~\cite{ionq_2020,ibm_2021, google_2020} -- so physical qubit requirements could easily rise to the megaqubyte scale for a $10^3$ lattice.

The generically dense $\hat{\mathsf{U}}(t)$ can only be efficiently constructed approximately. For the decomposition in noncommuting terms $\hat H = \sum_i \hat H_i$, a common approximation is \textit{trotterization}, whereby $\hat{\mathsf{U}}(t)\approx \mathcal U(t) =(\prod_i e^{-i \hat{H}_i\frac{t}{N}})^N$~\cite{trotter1959product,Suzuki:1985}.
%\equiv (e^{-i\hat{H}' \frac{t}{N}})^N$ with an approximate Hamiltonian $\hat{H}'$.
Implementing $\mathcal{U}(t)$ for a LGT may require large number of quantum gates to achieve desirable precision. 
%\YY{with high precisions in the trotterization step}. 
For example, in \cite{Kan:2021xfc} a $10^3$ lattice calculation of the shear viscosity $\eta$ in QCD with errors of $10^{-8}$ from trotterization and gate synthesis was estimated to require $\mathcal{O}(10^{49})$ T gates - the most expensive gate for error-correcting quantum computers. 
Though these estimates could be reduced by considering only the low-lying states \cite{Sahinoglu2020hamiltonian,Hatomura:2022yga} or by relaxing the precision requirement to the level of uncertainties from lattice truncation,
%finite lattice spacing and finite volume effects
gate costs are still expected to be inaccessible in the near-term.

Reducing quantum resources, either by implementing smarter quantum algorithms or performing classical processing, is thus strongly motivated. Gate reductions may be possible using other approximations of $\hat{\mathsf{U}}(t)$~\cite{PhysRevLett.123.070503,cirstoiu2020variational,gibbs2021longtime,yao2020adaptive,PhysRevLett.114.090502,Low2019hamiltonian}. At the cost of classical signal-to-noise problems, stochastic state preparation yields shallower circuits~\cite{Lamm:2018siq,Harmalkar:2020mpd,Gustafson:2020yfe,Yang:2021tbp}. Further, performing scale setting classically can reduce quantum resources~\cite{Osterwalder:1973dx,Osterwalder:1974tc,Carena:2021ltu}. LGT specific error correction or mitigation could also decrease costs~\cite{rajput2021quantum, Klco2021Hierarchy}.

In this \textit{letter}, we present a new direction for reducing quantum resources by using Hamiltonians with smaller discretization errors from finite differences. Quantum simulations can then be done at larger $a$, reducing the $\mathcal{O}((L/a)^d)$ qubits needed. 
We start with illustrating how to improve the commonly-used Kogut-Susskind Hamiltonian $H_{KS}$~\cite{PhysRevD.11.395} in the Symanzik improvement program~\cite{Symanzik:1983dc,Luscher:1984xn,Luscher:1985zq}, then derive time-evolution operators for the improved terms and construct the corresponding quantum circuits, followed by an explicit demonstration for $\mathbb{Z}_2$.

\begin{figure}[ht]
    \centering
    \includegraphics[width=0.8\linewidth]{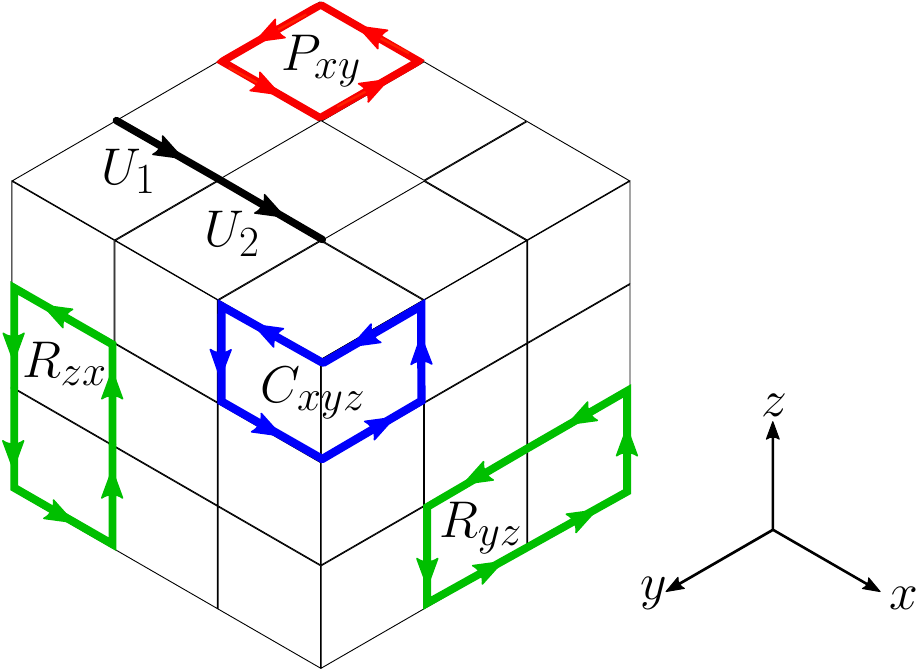}
    \caption{3d lattice with example contributions to $H_I$: the plaquette $P_{xy}$, rectangles $R_{yz}$ and $R_{zx}$, and the bent loop $C_{xyz}$, and the two links $U_1$ and $U_2$ used for $K_{2L}$.}
    \label{fig:latt}
\end{figure}

\textit{Improved Hamiltonians - }
For pure gauge theories, the classical Yang-Mills Hamiltonian can be written:
\begin{equation}
    H_{\rm co}=\frac{1}{2}\int \mathrm{d}^d x \tr\left[\mathbf{E}^2(\mathbf x)+ \mathbf{B}^2(\mathbf x)\right]
\end{equation}
where $\mathbf{E}(\mathbf x)$ and $\mathbf{B}(\mathbf x)$ are the electric and magnetic field strengths with spatial components $E_i(\mathbf x)$ and $B_i(\mathbf x)$. Alternatively, the magnetic energy density can be written in terms of $F_{ij}(\mathbf x)$, the spatial-spatial field strength tensor, as: $\frac{1}{2}\mathbf{B}^2(\mathbf x) = \frac{1}{2}\sum_{i<j} F^2_{ij}(\mathbf x)$ with Latin indices indicating spatial directions as shown in Fig.~\ref{fig:latt}.
%\wq{, for $d\neq3$, the $B$ field is not in spatial directions.}$F_{ij}(x) = \lambda^a F^a_{ij}(x)$ 
In terms of color components, $E_i(\mathbf x)=  E_i^b(\mathbf x) \lambda_b$, $B_i(\mathbf x)= B_i^b(\mathbf x) \lambda_b$, with $\lambda_b$ being generators of the gauge group. 
To ensure gauge invariance, lattice Hamiltonians
%\hank{missing an article} 
are built from gauge links $U_i(\mathbf x) = e^{i g a A_i(\mathbf x)}$ connecting lattice site $\mathbf x$ to its neighbor in the $i$ spatial direction, with $g$ being the gauge coupling and $A_i(\mathbf x)$ the lattice gauge field \cite{Wilson:confinement}. By replacing the magnetic field  $B_i(\mathbf x)$ term with the plaquettes $P_{ij}(\mathbf x)$ (see Fig.~\ref{fig:latt} for $i=x$ and $j=y$) built from $U_i(\mathbf x)$, and the electric field $E_i(\mathbf x)$ with the lattice electric field $L_i(\mathbf x)$, one arrives at $H_{KS}$~\cite{PhysRevD.11.395}:
\begin{align}
   H_{KS}&=K_{KS}+ V_{KS},\\
   K_{KS}=\sum_{\mathbf x, i} \frac{g_t^2}{a} \tr L^2_i&(\mathbf x),\,
   V_{KS}=-\sum_{\mathbf x, i<j} \frac{2}{g_s^2 a} \re\tr P_{ij}(\mathbf x)\notag.
\end{align}
As temporal and spatial directions are treated differently, coupling $g_t$ and $g_s$ are introduced for the kinetic term $K_{KS}$ and potential term $V_{KS}$, respectively. The discrepancy between $H_{KS}$ and $H_{\rm co}$ is of $\mathcal{O}(a^2)$, as seen by series-expanding $P_{ij}$ with $D_i$ denoting the covariant derivative:
\begin{equation}\label{eq:P_Taylor}
   P_{ij}=\mathbb{1}-\frac{g_s^2 a^4}{2}\bigg[F_{ij}^2+\frac{a^2}{12} F_{ij}(D_i^2+D_j^2) F_{ij}+\mathcal{O}(a^4)\bigg ].
\end{equation}

For Symanzik improvement, one adds terms to $H_{KS}$, and adjusts couplings to cancel the discretization errors~\cite{Luo:1998dx,Carlsson:2001wp}. The above classical $\mathcal{O}(a^2)$ error from $F_{ij}(D_i^2+D_j^2) F_{ij}$ can be cancelled by including the rectangle term $ R_{ij}(\mathbf x)$ (see Fig.~\ref{fig:latt}), as detailed in the Supplementary Material. 
At the quantum level $\mathcal{O}(g^2_s a^2)$ errors arise, requiring more terms, say the six-link bent loop terms $C_{ijk}(\mathbf x)$ (see Fig.~\ref{fig:latt}).

The improved Hamiltonian can be written as $H_{I}=K_I+V_{I}$ with the improved potential term $V_{I}=\beta_{V0}  V_{KS}+\beta_{V1} V_{\rm rect}+\beta_{V2} V_{\rm bent}$ and the improved kinetic term $K_I=\beta_{K0} K_{KS} + \beta_{K1} K_{2L}$~\cite{Carlsson:2001wp}. $V_{\rm rect}$ is defined as
\begin{equation}\label{eq:V1}
    V_{\rm rect}=\frac{2 }{a g_s^{2}} \sum_{\mathbf x, i<j}\re\tr\left[R_{i j}(\mathbf x) + R_{j i}(\mathbf x)\right], 
\end{equation} 
and $V_{\rm bent}$ has analogous expressions to $V_{\rm rect}$. To cancel the $\mathcal{O}(a^2)$ errors in $K_{KS}$, one adds the two-link term $K_{2L}$:
\begin{equation}\label{eq:improvedK}
    K_{2L}=\frac{g_t^2}{a} \sum_{\mathbf x, i} \tr \left[L_{i}(\mathbf x)  U_{i}( \mathbf x) L_{i}( \mathbf x+a \mathbf i) U_{i}^{\dagger}(\mathbf x)\right].
\end{equation}
For classical improvement, the couplings should be \cite{Luo:1998dx,Carlsson:2001wp}: $\beta_{V0}=\frac{5}{3}$, $\beta_{V1}=-\frac{1}{12}$, $\beta_{V2}=0$, $\beta_{K0}=\frac{5}{6}$ and $\beta_{K1}=\frac{1}{6}$. Perturbative improvements at the quantum level generate corrections of $\mathcal{O}(g^2a^2)$~\cite{Luscher:1985zq, Lepage_redesigningLQCD}. 
One can further nonperturbatively tune these couplings numerically. 
For quantum simulations, these couplings could be extracted via analytic continuation of Euclidean calculations~\cite{Carena:2021ltu}. The resulting $H_I$ then has leading errors of $\mathcal{O}(a^4)$ to $H_{\rm co}$.

Both $H_{KS}$ and $H_I$ can be derived from Euclidean actions via the transfer matrix in the continuous-time limit. The L\"uscher-Weisz action~\cite{Luscher:1984xn} was used to derive $H_I$~\cite{ Luo:1998dx,Carlsson:2001wp} and has improved errors of $\mathcal{O}(a^4)$ compared to the $\mathcal{O}(a^2)$ Wilson action used to derive $H_{KS}$ \cite{Creutz:1976ch}. For the L\"uscher-Weisz action, $a=0.4$ fm lattices were found to have similar discretization errors to $a=0.17$ fm lattices with the Wilson action~\cite{Alford:1995hw}.
Similar scaling is suggested by the limited direct studies of $H_I$ and $H_{KS}$~\cite{Carlsson:2003rf}.
As the number of qubits required is $\mathcal{O}((L/a)^d)$, using $H_I$ may require $\gtrsim 2^d$ fewer qubits in realistic quantum simulations for a fixed discretization error compared to $H_{KS}$. While we occupy ourselves with pure gauge theory, future effort should consider the $\mathcal{O}(a)$ fermion Hamiltonians~\cite{Spitz:2018eps} -- particularly for chiral fermions.

\textit{Circuit Design - }For quantum field theory calculations, $H_{I}$ is quantized by promoting the fields to operators: $U_i\to \hat U_i$, $L_i \to \hat L_i$.
The magnetic field basis is the eigenbasis of the link operator $\hat U$ while its Fourier transformation gives the electric field basis $\ket{L_i}$ diagonalizing $\hat L^2_i$.  
The quantum state of a link $\ket{U_i}$ is stored in a set of qubits - a link register.
Any gauge circuit can be built from a set of primitive gates~\cite{Lamm:2019bik} acting on link registers:

\begin{itemize}
    \item inverse gate: $\mathfrak{U}_{-1}|U_i\rangle=\left|U^{-1}_i\right\rangle$,
    \item left and right multiplication gates: $\mathfrak{U}_{\times}^L|U_i\rangle|U_j\rangle=|U_i\rangle|U_i U_j\rangle$, $\mathfrak{U}_{\times}^R|U_i\rangle|U_j\rangle=|U_i\rangle| U_jU_i\rangle$,
    \item trace gate: $\mathfrak{U}_{\operatorname{Tr}}(\theta)|U_i\rangle=e^{i \theta \operatorname{Re} \operatorname{Tr} U_i}|U_i\rangle$,
    \item Fourier gate: $
\mathfrak{U}_{F}\sum_{U_i}f(U_i)\left|U_i\right> = \sum_{L_i}\hat{f}(L_i)\left|L_i\right>$ with $\hat f$ denoting the Fourier transform of $f$.
\item L-phase gate: $\mathfrak{U}_{\operatorname{phase}}(\theta)$ is a gauge group specific phase rotation, implemented by a diagonal matrix.
\end{itemize}

We implement the quantum circuits for $\hat H_I$ term by term. Optimal quantum circuits depend on the underlying architecture -- in particular connectivity. We assume register connectivity between a pair of links sharing a common site (linear register connectivity). 

\begin{figure}[h]
\begin{subfigure}[t]{0.45\textwidth}
\includegraphics[width=0.75\linewidth]{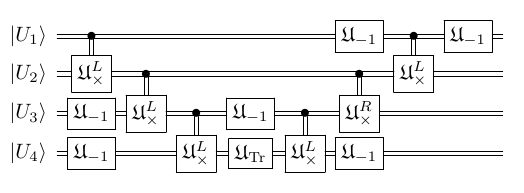}
\caption{$\mathcal{U}_{V_{KS}}$ assuming linear register connectivity.}
\label{fig:uv1}
\end{subfigure}\\
\begin{subfigure}[t]{0.48\textwidth}
\includegraphics[ width=0.9\linewidth]{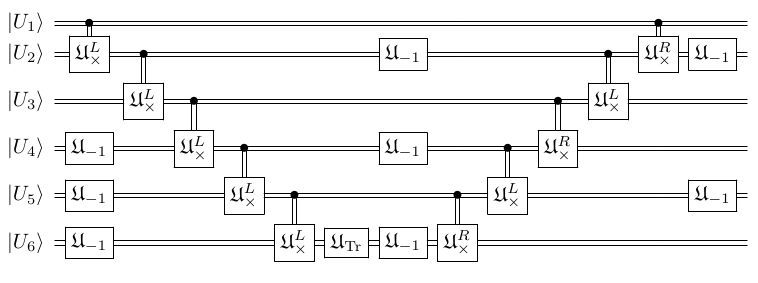}
\caption{$\mathcal{U}_{V_{\rm rect}}$ assuming linear register connectivity.}
\label{fig:uv12}
\end{subfigure}\\
\begin{subfigure}[t]{0.35\linewidth}
\raisebox{2mm}{\includegraphics[width=\linewidth]{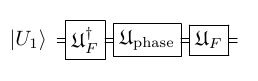}}
\caption{$\mathcal{U}_{K_{KS}}$.}
\label{fig:uk1}
\end{subfigure}
\begin{subfigure}[t]{0.62\linewidth}
\includegraphics[width=1\linewidth]{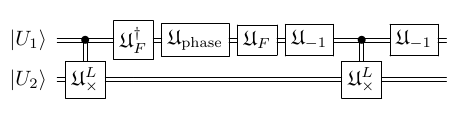}
\caption{$\mathcal{U}_{K_{2L}}$.}
\label{fig:UK12}
\end{subfigure}
\caption{Quantum circuits for the time evolution of $\hat{H}_I$.}
\end{figure}

$\hat V_I$ includes $\hat{P}_{ij}(\mathbf x)$ for every individual plaquette and the rectangles $\hat{R}_{ij}(\mathbf x)$ for every neighboring two plaquettes. We denote the circuits for $\hat V_{KS}$ as $\mathcal{U}_{V_{KS}}=e^{i\theta \re\tr\hat P_{ij}(\mathbf x)}$ (\fig{uv1}) and for the rectangles $\mathcal{U}_{V_{\rm rect}}=e^{i\theta \re\tr\hat R_{ij}(\mathbf x)}$ (\fig{uv12}), with the coupling and trotter step encoded in $\theta$. The circuit of \fig{uv12} with registers appropriately changed implements $\mathcal{U}_{V_{\rm bent}}$. 

The circuits $\mathcal{U}_{K_{KS}}=e^{i\theta \tr\hat{L}_1^2}$ for $\hat K_{KS}$ can be implemented by the L-phase gate $\mathfrak{U}_{\rm phase}$ in the electric field basis~\cite{Lamm:2019bik}, as shown in \fig{uk1}. To avoid dealing with $\hat L$ and $\hat U$ operators simultaneously, we rewrite $\hat K_{2L}$ as \begin{align}\label{eq:K1simplified}
   \hat K_{2L}=\frac{g^2_t}{a} \sum_{x, i} \tr [\hat R_{i}(\mathbf x)  \hat L_{i}( \mathbf x+a \mathbf i) ],
\end{align} 
using the right electric field operator~\cite{Zohar:2015hwa}:
\begin{align}\label{eq:R_def}
  & \hat R_i(\mathbf x)\equiv \hat U_{i}^{\dagger}(\mathbf x) \hat L_{i}(\mathbf x) \hat U_{i}(\mathbf x)  =\hat R_i ^b (\mathbf x)\lambda_b,
\end{align}
For simplicity, we denote the two succeeding links in one direction as $U_1$ and $U_2$ following Fig.~\ref{fig:latt}, and thus $\tr [\hat R_{i}(\mathbf x)  \hat L_{i}(\mathbf x+a \mathbf i) ]$ becomes $ \tr[\hat R_1 \hat L_2]$. For non-Abelian gauge theories, this sum of non-commuting terms ($ \hat R^b_1 \hat L^b_2$ with color index $b$) is difficult to implement.
We bypass this obstacle by decomposing $\hat R_1  \hat L_2$ as
\begin{equation}
     \tr ( \hat R_1  \hat L_2)=\tr [\hat L_2^2+\hat R_1^2-(\hat L_2-\hat R_1)^2]/2.
\end{equation}
With $\hat R^2 = \hat L^2$, the first two terms can be absorbed into $\hat{K}_{KS}$. Thus, for $\hat{K}_I$ the only new term is $\tr[(\hat L_2-\hat R_1)^2]$.
Defining the evolution operator $\mathcal{U}_{K_{2L}}\equiv e^{i\theta \tr(\hat{L}_2-\hat{R}_1)^2}$, and using $[\mathcal{U}_{K_{2L}},\hat{U}_1\hat{U}_2] = 0$, 
the matrix elements of $\mathcal{U}_{K_{2L}}$ are found to be (see Supplementary Material):
\begin{equation}
   \scalebox{0.95}[1]{$\bra {U_1',U_2'} \mathcal{U}_{K_{2L}}\ket {U_1,U_2}=\delta_{U_1'U_2', U_1U_2}\bra{U_1'}e^{i\theta \tr\hat{L}_1^2}\ket{U_1}$}, 
   \label{eq:step1}
\end{equation}
 The circuit in \fig{UK12} implements Eq.~(\ref{eq:step1}) by first storing the conserved quantity $U_1U_2$ in the second register $|U_2\rangle$ via $\mathcal{U}^L_{\times}$, then performing $e^{i\theta \tr \hat L_1^2}$ on $|U_1\rangle$ with the sequence $\mathcal{U}_{F}^\dagger\mathcal{U}_{\rm phase}\mathcal{U}_{F}$. Finally, we ensure the conserved product of $U_1 U_2$ imposed by $\delta(U_1'U_2'-U_1U_2)$ using the information stored in the second register via $\mathcal{U}_{-1}\mathcal{U}_{\times}\mathcal{U}_{-1}$. 
%\begin{figure}[h]
%\begin{center}
%\begin{subfigure}[t]{0.35\linewidth}
%\raisebox{2mm}{\includegraphics[width=\linewidth]{uk1}}
%\caption{$\mathcal{U}_{K_{KS}}$.}
%\label{fig:uk1}
%\end{subfigure}
%\begin{subfigure}[t]{0.62\linewidth}
%\includegraphics[width=1\linewidth]{UK_12}
%\caption{$\mathcal{U}_{K_{2L}}$.}
%\label{fig:UK12}
%\end{subfigure}
%\caption{Quantum circuits for the time evolution of $\hat{K}_I$.}
%\end{center}
%\end{figure}

While using $\hat{H}_I$ should require $\gtrsim2^d$ times fewer qubits, it requires additional gates to implement evolutions with the improved terms. Since the dominant quantum errors today are from decoherence and the entangling gates with error rates of $\mathcal{O}(10^{-2})$~\cite{wei2021quantum,lao2021software,Howe:2021mfd}, this increased cost may diminish the gain from using $\hat{H}_I$. We list the gate costs in terms of primitive gates in \tab{gateperlink} for one trotter step using either $\hat{H}_{KS}$ or $\hat{H}_{I}$.
Depending on which primitive gates dominate the circuits, the gate cost for $\hat{H}_I$ is 2 to 4 times that of $\hat H_{KS}$ per link register. For the group $\mathbb{Z}_{N}$ and $D_N$~\cite{Alam:2021uuq}, different primitive gates take approximately the same order of entangling native gates. Since $\hat{H}_I$ should require $\gtrsim 2^d$ fewer link registers, for the cases of $d=2,3$ we anticipate the same or fewer total primitive gate cost.

\begin{table}[ht]
    \begin{tabular}{c|c|c|c}
    \hline\hline
    Gate &$N[\hat{K}_{KS}+\hat{V}_{KS}]$&$N[\hat K_{\rm 2L}+\hat V_{\rm rect}]$& $\mathbb{Z}_2$ Impl.\\
    \hline
    $\mathfrak U_F$ &2     & 2& $H$\\
    \hline
    $\mathfrak U_{\rm phase}$&1    & 1& $R_z(\theta)$\\
    \hline
    $\mathfrak U_{\rm Tr}$&$\frac{d-1}{2}$& $d-1$& $R_z(\theta)$\\
    \hline
    $\mathfrak U_{-1}$& $3(d-1)$&$2+ 8(d-1)$ &$\mathbb{1}$\\
    \hline
    $\mathfrak U_{\times}$&$6(d-1)$ & $4+20(d-1)$&CNOT\\
    \hline
    \end{tabular}
    \caption{Number of primitive gates per link register per trotter step neglecting boundary effects (columns 2-3), implementation for $\mathbb{Z}_2$ (last column).}
    \label{tab:gateperlink}
\end{table}

\textit{Demonstration -}
\begin{figure*}[ht]
\begin{subfigure}[t]{0.4\textwidth}
\includegraphics[width=1\linewidth]{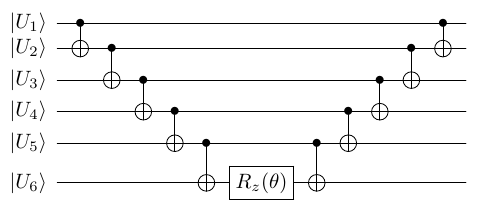}
\caption{}
\label{fig:Z2improvedHV}
\end{subfigure}
\begin{subfigure}[t]{0.25\textwidth}
\raisebox{10mm}{\includegraphics[ width=1\linewidth]{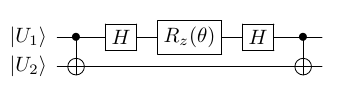}}
\caption{}
\label{fig:Z2improvedHK}
\end{subfigure}
\begin{subfigure}[t]{0.2\textwidth}
\includegraphics[width=\linewidth]{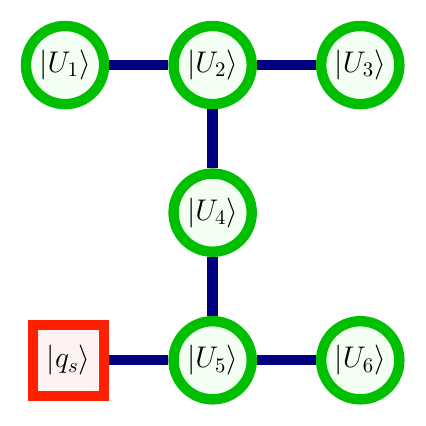}
\caption{}
\label{fig:connectivity}
\end{subfigure}
\caption{$\mathcal{U}_{V_{\rm rect}}$ (a) and $\mathcal{U}_{K_{2L}}$ (b) for $\mathbb{Z}_2$ gauge theory. (c) Link-to-qubit map on \texttt{ibm\_perth}.}
\label{}
\end{figure*}
For $\mathbb{Z}_2$ gauge theory, $\hat{H}_{I}$ can be mapped to Pauli matrices.  Choosing the magnetic field basis, the qubit state $\ket{0}$ ($\ket{1}$) represents the element 1 (-1) of $\mathbb{Z}_2$. 
Implementations of the primitive gates are listed in the last column of \tab{gateperlink}. We consider the most expensive $\mathbb{Z}_2$ gate, $\mathcal{U}_{V_{\rm rect}}$ on the 7-qubit \texttt{ibm\_perth} device (\fig{connectivity}). The connectivity of \texttt{ibm\_perth} prevents implementing $\mathcal{U}_{V_{\rm rect}}$ as shown in \fig{Z2improvedHV}. With the mapping from links to qubits shown in \fig{connectivity}, a transpiled version of the circuit with 12 CNOTs and 20 additional one-qubit gates are used. We use the benchmark value $\theta = \delta t/ (g_s g_t)= 0.811411$, precluding circuit optimization when using $\theta$ values such as $\pi/2$.

\begin{figure}[ht]
\includegraphics[width=0.85\linewidth]{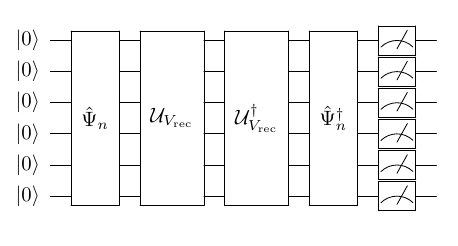}
\caption{$\mathcal{U}_{\mathrm{circ}}^{\ket{n}}$ for studying the errors of $\mathcal{U}_{V_{\rm rect}}$.}
\label{fig:ucirc}
\end{figure}

To quantify quantum errors, we evolve states with $\mathcal{U}_{V_{\rm rect}}$ and its inverse, and compare the measurement with noiseless expectations, implemented as the circuit $\mathcal{U}_{\mathrm{circ}}^{\ket{n}}$ in  \fig{ucirc}.
Without noise, the state preparation $\hat{\Psi}_n$ and $\mathcal{U}_{V_{\rm rect}}$ are exactly cancelled by their complex conjugations, thus all measurements return $\ket{0}^{\otimes6}$, and the distribution $P(w_H)$ of the Hamming weight $w_H$ -- the number of qubits measured in the $|1\rangle$ state -- returns $P(w_H)=\delta_{w_H, 0}$. In the noise-dominated limit where all states are equally populated, $P(w_H)$ follows the binomial distribution with 6 trials.
We take $\mathcal{F}_{\mathrm{rect}}^{|n\rangle}\equiv \sqrt{P(w_H=0)}$ as a definition of the quantum fidelity of $\mathcal{U}_{V_{\rm rect}}$ for the state $\ket{\Psi_n}=\hat{\Psi}_n \ket{0}^{\otimes6}$.
Determining the fidelity requires testing all the possible states $|\Psi_n\rangle$, a prohibitively expensive task~\cite{Chuang:1996hw}. Therefore we consider a restricted set consisting of  $\hat\Psi_{n}=\prod_{m\leq n} H_m^{\otimes}$ for $n \in [0,6]$ with $m$ indicating the qubit to which $H$ is applied. 

To mitigate the coherent noise dominating the CNOT errors, we implement Pauli twirling~\cite{Erhard_2019,li2017efficient, 2018efficienttwirling, 2013efficienttwirl, 2016efficienttwirling} which converts coherent errors into random errors in Pauli channels and has found success in low-dimensional lattice field theories~\cite{Yeter-Aydeniz:2022vuy}. The circuits are modified by wrapping each CNOT with a set of Pauli gates $\{\mathbb{1},X,Y,Z\}$ randomly sampled from sets satisfying
\begin{equation}
    \label{eq:paulicomp}
    \scalebox{0.95}[1]{$\left(\prod_{i}(\sigma^{b_i}_i)^{\otimes}\right)\text{CNOT}\otimes\mathbb{1}_4 \left(\prod_{i} (\sigma^{a_i}_i)^{\otimes}\right) = \text{CNOT}\otimes\mathbb{1}_4$}, 
\end{equation}
where the $i$-th qubit (including spectators) was rotated by $\sigma_i^{a_i}$ before the CNOT and by $\sigma_i^{b_i}$ after.
Despite the enormous number of possible circuits, prior work has found $\mathcal{O}(10)$ circuits to be sufficient for error mitigation~\cite{Erhard_2019}. Therefore we implemented 15 unique circuits and run each circuit $2^{13}$ times. We also compute $\mathcal{F}_{\mathrm{rect}}^{|6\rangle}$ without Pauli twirling to gauge its effect.

\begin{figure}
    \centering
    \includegraphics[width=\linewidth]{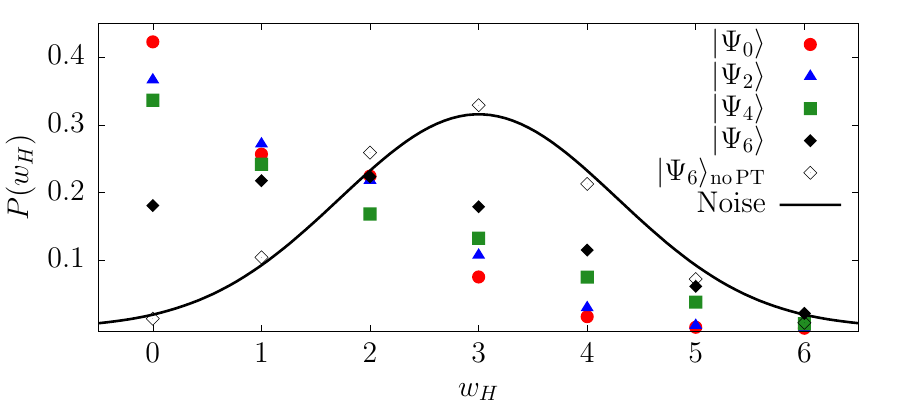}
    \caption{Probability of measuring Hamming weights for selected $|\Psi_n\rangle$ compared to the noise-dominated results. 
    In the noiseless limit, $P(w_H)=\delta_{w_H,0}$ for all $|\Psi_n\rangle$.
    %since the only $|0\rangle^{\otimes 6}$ should be measured.
    } 
    \label{fig:sim}
\end{figure}

With the above setup, we obtain the distribution $P(w_H)$ in Fig.~\ref{fig:sim} for selected $|\Psi_n\rangle$ and the state-dependent fidelities $\mathcal{F}_{\mathrm{rect}}^{|n\rangle}$ (Table~\ref{tab:qf}), yielding an average $\mathcal{F}_{\mathrm{rect}}=0.550$. Without Pauli twirling for $n=6$, $P(w_H)$ is indistinguishable from the noise-dominated limit while all the Pauli-twirled results are skewed toward the noiseless result, with states of lower $n$ (and consequently less average entanglement) being closer to the desired value. Comparing the results for $|\Psi_6\rangle$ with and without Pauli twirling we observe a fourfold improvement in fidelity -- clearly demonstrating the advantage from this error mitigation.

\begin{table}[ht]
        \begin{tabular}{c|c c c c c c c |c}
    \hline\hline
    $|\Psi_n\rangle$&
    $|\Psi_0\rangle$&$|\Psi_1\rangle$&$|\Psi_2\rangle$&$|\Psi_3\rangle$&$|\Psi_4\rangle$&$|\Psi_5\rangle$&$|\Psi_6\rangle$&$|\Psi_6\rangle_{\rm no\, PT}$\\\hline
    $\mathcal{F}_{\mathrm{rect}}^{|n\rangle}$&0.650&0.575&0.605&0.599&0.579&0.442&0.425&0.1194\\
    \hline
    \end{tabular}
    \caption{Measured state-dependent quantum fidelities with Pauli twirling and without it for $|\Psi_6\rangle_{\rm no\, PT}$.}
    \label{tab:qf}
\end{table}
For a single trotter step, the time evolution of $\hat{H}_{I}$ for a two-plaquette lattice with open boundary conditions requires at least 28 CNOTs (40 one-qubit gates): 12 CNOTs (20 one-qubit gates) for $\mathcal{U}_{V_{\rm rect}}$, at least 12 CNOTs (2 one-qubit gates) for the two $\mathcal{U}_{V_{KS}}$ and 4 CNOTs (6 one-qubit gates) for the two $\mathcal{U}_{K_{2L}}$, alongwith 12 one-qubit gates for $\mathcal{U}_{K_{KS}}$.  Assuming that the average fidelity depends on the total number of CNOT gates, we can estimate the single-trotter-step fidelity for $\hat H_I$: $\mathcal{F}_{\delta}\lesssim (\mathcal{F}_{\mathrm{rect}})^{28/12} \approx 0.25$.
Thus current devices are inadequate for real-time computations. However given the expected hardware improvements in the coming years \cite{ionq_2020,ibm_2021, google_2020}, 
$\mathcal{F}_{\delta}$ will be improved, allowing simulations of a two-plaquette lattice for $\mathbb{Z}_2$ gauge theory and direct comparisons between Hamiltonians. Alternatively, classical simulators could explore lattices up to $7^2$~\cite{Gustafson:2021jtq} to test improved Hamiltonians.

In this \textit{letter}, we designed the quantum circuits for simulating the improved Hamiltonian $\hat H_I$. Comparing to the commonly used $\hat H_{KS}$, $\hat H_{I}$ should allow quantum simulations with $\gtrsim 2^d$ fewer qubits.  With this reduction, we expect the gate count to be comparable or less than that of $\hat H_{KS}$ for theories with $d\geq 2$ despite increases of gate costs per link. For near-term numerical demonstrations, we constructed the circuits for $\hat{H}_{I}$ of the $\mathbb{Z}_2$ gauge theory and found that for \texttt{ibm\_perth} the fidelity of the 12 CNOT improved potential term is $\lesssim 0.550$. Our results suggest that alongside hardware developments, improved Hamiltonians can accelerate quantum simulations by years by reducing the number of qubits required, with optimistic prospects for $2+1d$ $\mathbb{Z}_2$ simulations in the near future. 
\begin{acknowledgments}
We would like to thank Erik Gustafson, Joseph Lykken and Michael Wagman for insightful discussions and comments on the manuscript. This work is supported by the Department of Energy through the Fermilab QuantiSED program in the area of ``Intersections of QIS and Theoretical Particle Physics". Fermilab is operated by Fermi Research Alliance, LLC under contract number DE-AC02-07CH11359 with the United States Department of Energy.  We acknowledge use of the IBM Q for this work. The views expressed are those of the authors and do not reflect the official policy or position of IBM or the IBM Q team.
\end{acknowledgments}
\bibliography{refs}% Produces the bibliography via BibTeX.
\end{document}